\documentclass[%
reprint,
superscriptaddress,
amsmath,amssymb,
aps,
]{revtex4-1}

\usepackage{color}
\usepackage{graphicx}
\usepackage{dcolumn}
\usepackage{bm}

\begin {document}

\title{Exact Results for First-Passage-Time Statistics in Biased Quenched Trap Models
}


\author{Takuma Akimoto}
\email{takuma@rs.tus.ac.jp}
\affiliation{%
  Department of Physics, Tokyo University of Science, Noda, Chiba 278-8510, Japan
}%


\author{Keiji Saito}
\affiliation{%
  Department of Physics, Keio University, Yokohama, 223-8522, Japan
}%


\date{\today}

\begin{abstract}
We provide exact results for the mean and  variance of  first-passage times (FPTs) of making a directed revolution  in the presence of a bias  
in heterogeneous quenched environments where the disorder is expressed by random traps on a ring 
with period $L$. FPT statistics are crucially affected by the disorder realization. In the large-$L$ limit, 
we obtain exact formulae for the FPT statistics, which are described by
 the sample mean and  variance for waiting times of periodically arranged traps. 
 Furthermore, we find that these formulae are still useful  for nonperiodic heterogeneous environments; i.e, 
the results are valid for almost all disorder realizations.
 Our findings are fundamentally important 
 for the application of FPT to estimate diffusivity of a heterogeneous environment under a bias.
\end{abstract}

\maketitle


\section{Introduction}
Encountering a reactive molecule or finding a reactive site by a molecule is the first step in chemical reactions. 
Therefore, finding a specific target in stochastic processes is a fundamental problem in the context of chemical 
as well as biological reactions \cite{redner2001}. In particular, this target-search problem attracts significant interests 
in biomolecular reactions in cells such as transcription factors searching for a specific DNA sequence \cite{Berg1981, von2007,mirny2008}. 
Many stochastic models have been utilized to unravel how biomolecules can efficiently reach the targets in cells 
\cite{loverdo2008, Mirny2009, lomholt2009, Benichou2011}, where a combination of 3D free diffusion and 1D sliding motion on DNA
plays a vital role in reducing the first-passage time (FPT) to the target. 

The 1D sliding motion is crucially affected by interactions between a searching molecule and DNA sequences \cite{shimamoto1999}. 
DNA sequences exhibit anomalous fluctuations, such as long correlations and $1/f$ fluctuations \cite{li1992}. Thus, 
a 1D sliding motion on DNA is described according to the diffusion in a quenched heterogeneous environment. 
In experiments, the diffusion coefficients of a repressor protein diffusing on DNA are obtained by single-particle-tracking measurements and 
show large trajectory-to-trajectory fluctuations \cite{Wang2006,Graneli2006}.
These fluctuations are evidence of the heterogeneity of the environment. In fact,  intrinsic fluctuations of the diffusion 
coefficients are observed in diffusion on heterogeneous environments such as the quenched trap model (QTM), which represents a random walk (RW)
on a random energy landscape, and an annealed model of the QTM, which presents continuous-time RW (CTRW)  \cite{He2008, Miyaguchi2011,*Miyaguchi2015}.

The FPT statistics in heterogeneous environments  are key quantities in target-search problems 
\cite{Slutsky2004PRE,slutsky2004}
and play an important role in estimating the diffusion coefficient \cite{zwanzig1988diffusion}.
  In higher dimensions, the CTRW provides a good description of diffusion in quenched heterogeneous environments, and hence the FPT statistics with the CTRW have intensively been studied \cite{rangarajan2000,Barkai2001, Condfamin2007,Krusemann2014,*krusemann2015}. 
  However, in 1D systems, the diffusion in quenched potentials cannot be approximated with the CTRW, and the FPT statistics is not well understood. It is known that 1D quenched systems exhibit distinct behaviors of diffusion \cite{bouchaud90,Miyaguchi2011,*Miyaguchi2015}.
  Hence, filling the lacuna on the FPT statistics in 1D systems is critical to update fundamental understanding of diffusion in 
  the quenched potential.

 In this paper, we clarify several properties of the FPT statistics inherent to the quenched potential by looking at the 1D biased QTM.
  We first consider a periodic random potential and derive exact FPT statistics. We next show that the formulae are available to understand the nonperiodic potential also. 
Although the periodic potential landscape is employed to simplify the setup in this paper, 
stochastic dynamics in the periodic potential have been intensively studied analytically \cite{Reimann2001, *Reimann2002,  dean2014} and also experimentally \cite{Reimann2008,Hayashi2015,Ma2015, *Ma2017, Kim2017}. Moreover, a bias in diffusion processes induces surprising phenomena such as giant acceleration of diffusivity in periodic potentials \cite{Reimann2001, *Reimann2002, Reimann2008}, field-induced superdiffusion \cite{shlesinger1974,margolin2002, berkowitz2006,Burioni2013, *Burioni2014}, and distinct initial ensemble dependence of diffusivity in disordered media such as the CTRW \cite{Akimoto2018ergodicity, hou2018biased}. Our analysis also unravels several indications of the effects caused by quenched disorder on these phenomena.


\begin{figure}
\includegraphics[width=1.\linewidth, angle=0]{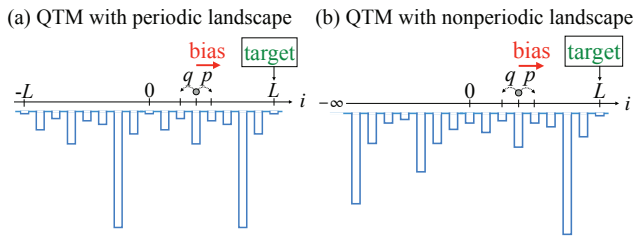}
\caption{Schematic representation of a biased QTM, where we represent a
random energy landscape as a 1D energy landscape
(a) with period $L$ ($L=8$) and (b) without periodicity. 
When a particle escapes from a valley of a random potential, it jumps to the right valley with probability $p$. 
A particle starts at the origin and the target is located at  site $L$. 
Note that there is no target on the left-hand side (no target at $0, -L, -2L, \cdots$).
In other words, we consider a time when a particle makes a directed (clockwise or counter-clockwise) revolution on a ring.
 }
\label{biased-qtm}
\end{figure}

\section{Model and Main Results}We consider  effects of bias on the FPT statistics in a quenched heterogeneous environment. 
In particular, we use a biased RW on a 1D quenched random energy landscape, which 
 is periodically arranged, i.e., a biased QTM with a periodic landscape (see Fig.~\ref{biased-qtm}(a)) \cite{bouchaud90}. 
 The target is located only at site $L$ only while the energy landscape is periodic.
Probabilities of the right and  left jumps are given by $p$ and $q=1-p$, respectively. 
A biased RW implies $p\ne 1/2$, and here we consider  $p>1/2$.
 We assume that the tops of the potentials are flat; i.e., the tops are the same height, implying that 
probability $p$ does not affect the shape of the random energy landscape.  
This physical situation is relevant to a biased diffusion in heterogeneous comb-like structures \cite{berezhkovskii2015biased}, e.g., porous media \cite{berkowitz2006} 
and neuronal dendrites \cite{mendez2013comb, jose2018trapping}, 
in which the bias is considered to be a flow in the backbone.


We assume that the lattice constant is set to unity and number $L$ of  lattice sites with different energies is finite ($L<\infty$).
At each site, depth $E~(>0)$ of the energy trap is randomly assigned. In particular, the depths are 
independent and identically distributed (IID) random variables with an exponential distribution,
$\rho(E) = T^{-1}_g \exp(-E/T_g)$, where $T_g$ is called the glass temperature. 
A particle can escape from a trap and jump to one of the nearest neighbors.  
A waiting time when a particle escapes from the $i$th trap is a random variable, and  the distribution follows the exponential distribution
with mean $\tau_i$: $\psi^{(i)}(\tau)=\tau_i^{-1} e^{-\tau/\tau_i}$ \cite{MELNIKOV1991}. 
Mean waiting time $\tau_{i}$ follows the Arrhenius law, i.e., $\tau_{i} \propto \exp(E_{i}/T)$, 
where $E_{i}$ is the depth of the energy at the $i$th trap and $T$ denotes the temperature.  
Through the Arrhenius law and the energy distribution $\rho (E)$, the probability density function  of  {$\tau_i$}
 follows a power law, i.e., $\psi_{\alpha} (\tau) \propto \tau^{-1 - \alpha }$ 
with $\alpha \equiv T/T_g$ \cite{Bardou2002,Akimoto2018}. When the temperature is below $T_g$, i.e., $\alpha <1$, the mean  of $\tau_i$
 diverges, inducing anomalous features such as anomalous diffusion and aging \cite{bouchaud90, Bertin2003, Miyaguchi2011, *Miyaguchi2015}.
Note that sample mean waiting time 
\begin{equation}
\mu_L = \frac{1}{L} \sum_{i=0}^{L-1} \tau_{i}
\end{equation}
for a fixed disorder in the QTM with a periodic landscape  never diverges when $L<\infty$. 

We consider the FPT, i.e., a time when a particle starting from the origin reaches the target (site $L$) for the first time. 
As the main results of this study, we show the mean FPT (MFPT) and the variance of the FPT (VFPT) for 
 a given quenched periodic landscape  for large $L$:
\begin{align}
 \langle T \rangle_L                 & \sim  \frac{ L \mu_L}{p-q} \equiv T_{\rm MFPT} \, ,   \label{mfpt_qtm} \\ 
 \langle \delta T^2 \rangle_L   & \sim  \frac{L \{ \sigma_L^2(p-q) + \mu_L^2\}}{(p-q)^3} \equiv T_{\rm VFPT} \, , 
 \label{var_fpt_qtm}
 \end{align}
where $\delta T \equiv T - \langle T \rangle_L $ and $\sigma_L^2$ is the sample variance, i.e.,
\begin{equation}
\sigma_L^2 = \frac{1}{L} \sum_{i=0}^{L-1} \tau_i^2 - \mu_L^2 \, .
\end{equation}
Sample variance $\sigma_L^2$ quantifies the degree of heterogeneity. The VFPT for $\alpha<1$ 
becomes $ \langle \delta T^2 \rangle_L   \sim L (\sigma^2_L + \mu_L^2)/(p-q)^2$ because  $\mu_L^2 =o( \sigma^2_L)$
 for large $L$. 
In this paper, we discuss physics behind the exact results and provide a brief sketch of the derivation.
We note that the MFPT diverges when $p=q$ because the mean return time to the origin diverges. 
Therefore, the results include case $p=q$. 
When the bias is small, $p-q$ can be expressed as $p-q\cong F/T$, where $F$ is an external field. 
In this situation,  the leading orders for small $F$ dependencies of the FPT statistics 
are represented as $\langle T \rangle_L\propto 1/F$ and $\langle \delta T^2 \rangle_L \propto 1/F^3$.  

A crucial aspect of the FPT statistics is that they are expressed by the statistics of the waiting times in the quenched heterogeneous environment. 
Note that the results do not explicitly include parameter $\alpha$. 
Instead, they depend on $\mu_L$ and $\sigma_L$, which are finite and depend on each realization of the disorder.
In addition, the VFPT are connected to the diffusivity in the biased QTM on a ring \cite{Reimann2001, *Reimann2002}. Therefore, the results play a significant role in estimating diffusivity, as discussed later. 

 The comparison of properties of FPTs in the QTM with those in the CTRW is intriguing.  In the CTRW 
 the waiting-time distribution is identical for all sites. Although the FPT distribution in CTRW has already been studied \cite{Condfamin2007}, the explicit forms of the FPT statistics have never been obtained so far. 
 Importantly, our results also lead to the exact expressions of the  biased CTRW  \cite{suppl}, which  
 are given by
\begin{align}
 \langle T_{\rm \tiny ctrw} \rangle_{L}  & \sim  \frac{  L\mu}{p-q}\,   , \label{mfptvar_fpt_ctrw}   \\
\langle \delta T_{\rm \tiny ctrw}^2 \rangle_L   &\sim  { L \left\{  \sigma^2 (p-q)^2 + 4pq \mu^2 \right\}\over (p-q)^3}  ,
\label{vfptvar_fpt_ctrw}  
\end{align}
where $\mu$ and $\sigma^2$ are the mean and the variance of the waiting-time distribution.  
The MFPT and VFPT  diverge for $\alpha <1$ and $\alpha <2$, respectively, 
while the QTM results are finite for all regimes of $\alpha$. 
The VFPT of the biased-CTRW is not given by a straightforward extension obtained from that of the biased QTM. This is unexpected because the CTRW is believed to be a good approximation of the QTM when a bias is added.

\begin{figure}
\includegraphics[width=.7\linewidth, angle=-90]{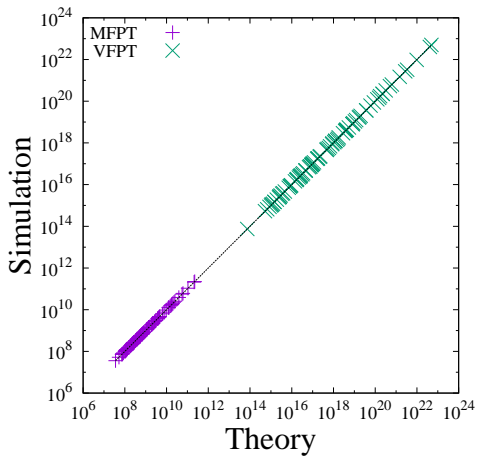}
\caption{ Correlation plot for the MFPT and  VFPT in the QTMs with periodic landscapes ($\alpha=0.5$ and $L=10^4$).  Symbols represent results of numerical simulations for 100 different 
disorder realizations. 
 }
\label{FPT-statistics-periodic}
\end{figure}

\section{Sample-to-sample fluctuations} We present numerical verifications of exact results~(\ref{mfpt_qtm}) and (\ref{var_fpt_qtm}) to see how these formula work for each disorder realization. Figure~\ref{FPT-statistics-periodic} shows the MFPT and VFPT for $100$ disorder realizations, where the numerical values are plotted as a function of the theoretical values. All the results are collapsed on the $y=x$ line, which shows a perfect agreement between the theory and numerical results. Note that the numerical results are provided for parameter $\alpha=0.5$, where the disorder realizations show large sample-to-sample fluctuations. The MFPT and VFPT also exhibit strong sample-to-sample fluctuations because the disorder strongly affects $\mu_L$ and $\sigma_L^2$. 
Nevertheless, the theoretical values for different realizations are remarkably correct. 

Next, we discuss sample-to-sample fluctuations by considering a disorder average. In general, when the heterogeneity of a disorder realization is sufficiently weak, physical observables comprise self-averaging (SA) properties \cite{bouchaud90}. Here, we quantify sample-to-sample fluctuations by the SA parameters defined as follows:
\begin{eqnarray}
{\rm SA}(L;  \langle \mathcal{O} \rangle_L) &\equiv& \frac{\langle \langle \mathcal{O} \rangle_L^2 \rangle_{\rm dis} - \langle \langle \mathcal{O} \rangle_L \rangle_{\rm dis}^2}
{\langle \langle \mathcal{O} \rangle_L \rangle_{\rm dis}^2},
\label{SA-def}
\end{eqnarray} 
where $\langle \cdot \rangle_{\rm dis}$ indicates the disorder average and observable $\mathcal{O}$ is $T$ or $\delta T^2$. The
vanishing of these quantities implies a perfect realization of SA, and hence
these parameters systematically quantify a degree of the SA property.
This definition is analogous to that of the diffusivity in the QTM discussed in \cite{Akimoto2016, *Akimoto2018}. 

As $\mu_L$ is a random variable, the SA parameter for the MFPT, i.e., $\mathcal{O}=T$, can be rewritten as
\begin{equation}
{\rm SA}(L;  \langle T \rangle_L) = \frac{\langle \mu_L^2 \rangle_{\rm dis} - \langle \mu_L \rangle_{\rm dis}^2}
{\langle \mu_L \rangle_{\rm dis}^2}= \frac{\langle \tau^2 \rangle_{\rm dis} - \langle \tau \rangle_{\rm dis}^2}
{L\langle \tau \rangle_{\rm dis}^2 } . \nonumber 
\label{SA_MFPT2}
\end{equation} 
For $\alpha <2$, the SA parameter is infinite because $\langle \tau^2 \rangle_{\rm dis}$ diverges, while 
for $\alpha >2$, it vanishes in the large-$L$ limit, implying that the SA is satisfied for $\alpha > 2$, while it is violated for $\alpha <2$.
Hence, the transition between SA and non-SA occurs at $\alpha_c=2$ for quantity $\langle T \rangle_L$.
Similarly, the transition from SA to non-SA for quantity $\langle \delta T^2 \rangle_L$ can be discussed through Eq.~(\ref{SA-def}). 
From a similar calculation, 
the critical value can be easily obtained as $ \alpha_c =4$. 
That is, the VFPT has an SA property for $\alpha > 4$, while it is broken for $\alpha <4$.

\begin{figure}
\includegraphics[width=1.0\linewidth, angle=0]{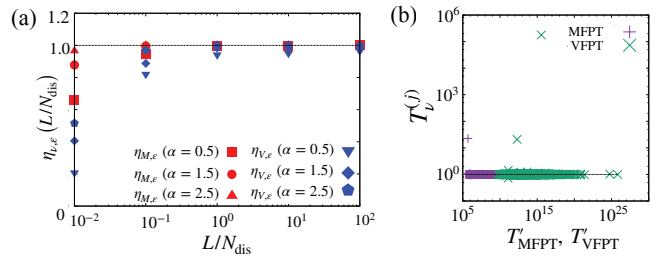}
\caption{(a) Ratio $\eta_{\nu,\varepsilon}$ as a function of $L/N_{\rm dis}$ ($N_{\rm dis}=10^3, \varepsilon=0.5$ 
and  $p=0.8$) for nonperiodic landscapes. 
Symbols represent results of numerical simulations. We used  $10^4$  thermal paths to calculate the MFPT or  VFPT for a fixed disorder realization.
(b) Correlation plot for the MFPT and  VFPT in the infinite 1D systems with nonperiodic landscapes ($\alpha=0.5$ and $L=10^3$). 
Numerical simulations of the MFPT and  VFPT for $10^3$ disorder realizations  are presented by symbols. 
 }
\label{FPT-infinite}
\end{figure}

\section{Numerical argument of the biased QTM with nonperiodic landscape} We discuss the FPT statistics in the biased QTM in which the potentials are arranged randomly in the infinite line (Fig.~\ref{biased-qtm}(b)).
 For a finite bias and large $L$, a particle will experience deep potentials mainly in the positive regime $i>0$.
Therefore, the FPT might be dominated by waiting times for  regime $i>0$. If this is true,
the exact results for the periodic QTM may still be useful to understand the FPT statistics in the infinite 1D systems.  
To discuss the validity of this theory, we define the following two quantities
\begin{equation}
 T_{\rm M} \equiv {\langle T \rangle_L\over T_{\rm MFPT}' }~ ~{\rm and}~~T_{\rm V} \equiv {\langle  \delta T^2 \rangle_L \over T_{\rm VFPT}' }, 
 \end{equation}
 where  $T_{\rm MFPT}'$ and $T_{\rm VFPT}'$ are the same expressions as in Eqs.~(\ref{mfpt_qtm}) and (\ref{var_fpt_qtm}), respectively. We should note that $\mu_L$ and $\sigma_L^2$ in these expressions are calculated from $\tau_i$ for $i=0, \cdots, L-1$ (in other words, we do not use the information of  potentials for $i<0$).  In addition, to quantify how our prediction works well, we introduce the following ratio:
\begin{equation}
\eta_{\nu, \varepsilon} \left( \frac{L}{N_{\rm dis}} \right) \equiv \frac{ 1}{N_{\rm dis}}  \sum_{j=1}^{N_{\rm dis}}  I_{[1-\varepsilon,1+\varepsilon]} ( T_\nu^{(j)}) ,   ~~{\nu={\rm M}, {\rm V}} 
 \end{equation}
where $T_{\nu}^{(j)}$ is a numerical value of $T_{\nu}$ for the $j$th realization of disorder and $I_A(x)$ is an indicator function, i.e., $I_A(x)=1$ if $x\in A$ and 
$I_A(x)=0$ otherwise. This quantifies a ratio that $T_{\nu}^{(j)}$ is within the corresponding theory with a $\varepsilon$-dependent  accuracy.   
As shown in  Fig.~\ref{FPT-infinite}(a),  the ratio approaches to 1 with increasing $L$. 

To understand more details at the level of each disorder realization, we next consider each $T_{\nu}^{(j)}$. In Fig.~\ref{FPT-infinite}(b), we present numerical data of $T_{\nu}^{(j)}$as a function of $T_{\rm MFPT}'$ or $T_{\rm VFPT}'$ depending on $\nu={\rm M}$ or ${\rm V}$, respectively. 
Figure~\ref{FPT-infinite}(b) shows that $T_{\rm MFPT}'$ and $T_{\rm VFPT}'$ are very good approximations of $\langle T \rangle_L$ and $\langle \delta T^2 \rangle_L$ for almost all realizations, except for small number of realizations with extremely large deviations.  
In such rare samples, significantly large waiting times are assigned for $i<0$ and small $|i|$
(see \cite{suppl}). Except for such rare samples, the biased-QTM results with a periodic landscape are surprisingly useful 
in nonperiodic landscapes.

\section{Derivation of Main results}
We now briefly describe the derivation of our results.  
We divide our explanation into two steps: step 1 explains about the FPT in the standard biased RW (without random traps), and step 2 explains about
 the FPT in the biased QTMs with periodic landscapes. 

\subsection{Step 1: Statistics of the numbers of visits }

In the first step, we outline our main strategy to derive the main results for the QTM, which gives us  another derivation for  known results of the FPT statistics in the classical RW. 
  The main strategy is to use statistics of the number of visits at each site. 
  A similar quantity has also been employed to study diffusion of nonbiased motions \cite{bouchaud90, Burov2011}. A biased RW was studied in the context of the classical ruin problems \cite{Feller1968}. 
In the ruin problems, a gambler with a capital wins or loses a dollar with probabilities $p$ or $q$, respectively.  
The FPT from the origin to $L$ site in the RW, i.e., $T_{\rm rw}$, correspond to a duration of the game in the ruin problems, in which
 the game is over when one of the two players is ruined. Here, we consider that one of the players has infinite capital and his win 
probability is $p>1/2$.
The generating function for the FPT in the classical ruin problems was derived in \cite{Feller1968}, and the MFPT and  VFPT  are respectively obtained as   
\begin{equation}
\langle T_{\rm rw}  \rangle_L = \frac{L}{p-q}
~~{\rm and}~~
\langle \delta T_{\rm rw}^2 \rangle_L  =  \frac{4pqL}{(p-q)^3}.
\label{MFPT_RWvar_rw}
\end{equation} 
While  these results are exact for any $L$, we will consider the large-$L$ limit to derive the biased-QTM results.
Note that $\langle T_{\rm rw} \rangle$ is a special case of Eq.~(\ref{mfpt_qtm}), while $\langle \delta T_{\rm rw}^2 \rangle_L$ is not 
a special case of Eq.~(\ref{var_fpt_qtm}) (see Eq.~(\ref{var_fpt_qtm2}) for the general result, which reproduces $\langle \delta T_{\rm rw}^2 \rangle_L$). 

The FPT from the origin to $L$ site can be represented by the sum of the numbers of  visits at each site, i.e., 
\begin{equation}
T_{\rm rw} = \sum_{i=0}^{L-1} k_i, 
\end{equation}
where $k_i$ is the number of visits to the $i$th site until the particle reaches  site $L$. 
Note that $k_i$ includes the number of the visits at sites $+i-nL$ ($n=1, 2, \cdots$).  
To obtain the moments and  correlation function of $k_i$, we consider the large-$L$ limit. In this limit, the probability that a particle reaches  site $-L$ becomes zero; i.e., a particle never visits the site $-L$. 
In the large-$L$ limit, one can obtain the generating function of $k_i$ until the particle reaches  site $L$ (see Appendix~A):
\begin{equation}
Z(\lambda) 
\to \frac{p -q}{e^{-\lambda}-2q},
\end{equation}
which yields the followings: 
\begin{equation}
\langle k_i \rangle \sim 1/(p-q) ~~{\rm and}~~ 
\langle k_i^2 \rangle - \langle k_i \rangle^2 \sim 2q/(p-q)^2 
\end{equation} 
for $L\to\infty$.
Moreover, correlation $\langle k_i k_j \rangle$ can be obtained exactly  as follows:
\begin{equation}
C_l \equiv \langle k_i k_{i+l} \rangle - \langle k \rangle^2 
\to \frac{\varepsilon^l}{(p-q)^2} \quad (L\to \infty),
\end{equation}
where $\varepsilon = q/p$, $\langle k \rangle = 1/(p-q)$, and $l \equiv |i-j|$. This correlation is derived in Appendix~B.
Note that $C_l$  does not depend on $i$, and $l$ is an arbitrary integer satisfying $l < L -i$. 
The MFPT and  second moment of the FPT can be respectively represented as 
\begin{equation*}
\langle T_{\rm rw} \rangle_L \sim L \langle k \rangle ~~{\rm and}~~ 
\langle T_{\rm rw}^2 \rangle_L \sim L (\langle \delta k^2 \rangle + \langle k \rangle^2)  + \sum_{i\ne j} \langle k_i k_j \rangle, 
\end{equation*}
where $\langle \delta k^2 \rangle = 2q/(p-q)^2$. 
Thus,  the VFPT becomes 
\begin{align}
\langle \delta T_{\rm rw}^2 \rangle_L  & = L \langle \delta k^2 \rangle   +
2\sum_{l=1}^{L-1} (L-l)C_l \sim  L (\langle \delta k^2 \rangle  + 2C) \, ,
\end{align} 
where $C = \sum_{l=1}^{L-1} C_l$. Note that  $\sum_{l=1}^{L-1} lC_l$ converges to a constant for $L\to\infty$ because 
$C_l$ decays exponentially to zero. 
In the large-$L$ limit, we have
\begin{equation}
C 
=\frac{q}{(p-q)^3}.
\end{equation}
This leads to the desired results of the MFPT and   VFPT.
Note that the RW results of the FPT statistics are exact for any $L$ \cite{Feller1968}. 

\subsection{Step 2: Derivation of the QTM results}
By using the same technique used in step 1, the biased-QTM results can be derived. 
Note that the FPT in the QTM can be obtained by 
\begin{equation}
T=\sum_{i=0}^{L-1} T_i\quad {\rm with}\quad
 T_i = \sum_{l=1}^{k_i} \tau_m^{(i)}, 
 \label{fpt_ki}
 \end{equation}
where $\tau_m^{(i)}$ is the waiting time for the $m$th visit to
   site $i$ and $T_i$ is  the occupation time at  site $i$. 
 The mean of $T_i$ can be calculated as
 $
 \langle T_i \rangle = \langle \tau_1^{(i)} + \cdots + \tau_{k_i}^{(i)}\rangle = \langle k \rangle \tau_i.
 $
Thus, the MFPT in the large-$L$ limit is given by
 $\langle T \rangle_L \sim \langle k \rangle \sum_{i=0}^{L-1} \tau_i \sim L \mu_L/ (p-q)$, i.e, Eq.~(\ref{mfpt_qtm}).
 This is a simple extension of the MFPT for a biased RW and is easily obtained by multiplying 
 $\langle T_{\rm rw}  \rangle_L$ by $\mu_L$. 

 Using $\langle T_i^2 \rangle$, we also have  the VFPT in the biased QTM (the details are given in Appendix~C). 
 In the large-$L$ limit,  the VFPT for $\alpha>1$  becomes
 \begin{equation}
\frac{ \langle \delta T^2 \rangle_L}{L}   \sim  \frac{2q (\sigma_L^2 + \mu_L^2)}{(p-q)^2}  +\frac{\langle \delta \tau^2 \rangle_L }{p-q}+  \frac{2q \mu_L^2}{(p-q)^3},
 \label{var_fpt_qtm2}
 \end{equation}
 where 
 \begin{equation}
 \langle \delta \tau^2 \rangle_L = \frac{1}{L}\sum_{i=0}^{L-1} (\langle (\tau_l^{(i)})^2 \rangle -\langle \tau_l^{(i)} \rangle^2).
 \end{equation} 
 For $\alpha<1$, $\mu_L^2$ can be ignored because  $\mu_L^2=o(\sigma_L^2)$.
  In the QTM, $\langle \delta \tau^2 \rangle_L =\sigma_L^2 + \mu_L^2$, which  gives our main claim, i.e., Eq.~(\ref{var_fpt_qtm}). 
 Note that Eq.~(\ref{var_fpt_qtm2}) is a more general expression of  the VFPT than Eq.~(\ref{var_fpt_qtm}), 
 which includes the VFPT in the classical RW, $\langle \delta T_{\rm rw}^2 \rangle_L$. 
Moreover, it is straightforward to derive exact results for the biased CTRW. In the CTRW, the waiting-time distribution is identical for all sites. Thus, sample variance $\sigma_L^2$ in CTRWs is zero. 
Replacing $\mu_L$ and $\langle \delta \tau^2 \rangle_L$ with  $\mu$ and $\sigma^2$ gives
the exact expressions, i.e., Eqs.~(\ref{mfptvar_fpt_ctrw}) and (\ref{vfptvar_fpt_ctrw}). 

\section{discussion}
We derived  the MFPT and  VFPT in the QTM with a random periodic potential  in the presence of bias. 
In the large-$L$ limit, our formulae 
provide the exact expressions of the FPT statistics in the biased CTRW. Unexpectedly,  the VFPT values of the biased CTRW and  QTM are distinctive. 
Furthermore, the results for the biased QTMs with periodic landscapes are still surprisingly useful even when the energy landscape is not periodically arranged  in the 1D line. 

Finally, we briefly discuss the diffusion coefficient in the biased QTM on a ring.
Here, we apply our formulae to diffusion in the system with period $L$. 
Let $n_t$ be the number of events in which a particle makes a directed revolution (biased direction).
As the time intervals between the events are IID random variables, 
the process of $n_t$ is described as a renewal process.
By renewal theory \cite{Cox}, the mean of $n_t$ is given by
$\langle n_t \rangle = t/\langle T \rangle_L$. 
Displacement $\delta x_t$ is represented by
$\delta x_t = L n_t + C_L$, 
where $C_L$ is a random variable, and the mean has the order of $L$, i.e., $\langle C_L \rangle =O(L)$.
Thus, $\langle \delta x_t \rangle$ becomes 
\begin{equation}
\langle \delta x_t \rangle \sim L \langle n_t \rangle = \frac{(p-q)t}{\mu_L} \quad (t\to\infty).
\end{equation}
Moreover, 
using the variance of $n_t$ \cite{Cox}, 
we have
\begin{equation}
\langle \delta x_t^2 \rangle -\langle \delta x_t \rangle^2 \sim 
 \left(  \frac{ (p-q) (\sigma_L^2 +\mu_L^2)}{   \mu_L^3} + \frac{2q}{\mu_L} \right) t.
\label{var_xt}
\end{equation}

\section*{ Acknowledgement} We thank E. Barkai and T. Miyaguchi for giving us fruitful comments.
This work was supported by JSPS KAKENHI Grant Number 16KT0021, 18K03468 (TA), and JP17K05587 (KS).

%
\appendix

\section{Details for a biased random walk}
A biased RW was studied in the context of the classical ruin problems \cite{Feller1968} in which a gambler with a finite/infinite capital wins or loses a dollar with probabilities $p$ and $q$, respectively. Let us consider the probability of his ruin when the initial capital is $z$ and the other player's capital is $L-z$. In the language of the RW, this probability corresponds to the probability that a particle starting from  site $z$ ($0 \leq z \leq L$) 
reaches  site $0$ without visiting  site $L$. This probability denoted by $q_z^L$ is known as \cite{Feller1968}
\begin{equation}
q_z^L = \dfrac{\varepsilon^L - \varepsilon^z}{\varepsilon^L - 1}, 
\end{equation}
where $\varepsilon = q/p$. Moreover, the probability of his win is given by $p_z^L = 1 - q_z^L$ because the game will end in the future with 
probability 1. 

Here, we assume $p>1/2$ and the random walker starts at the origin. 
The FPT from the origin to $L$ site can be represented by the sum of the numbers of the visits at each site, i.e., 
\begin{equation}
T_{\rm rw} = \sum_{i=0}^{L-1} k_i, 
\end{equation}
where $k_i$ is the number of visits to the $i$th site until the particle reaches the site $L$. 
We note that $k_i$ includes the number of the visits at sites  $+i-nL$ ($n=1, 2,  \cdots$). 

To obtain the moments and the correlation function of $k_i$, we consider the large-$L$ limit. 
In this limit, the probability that a particle starting at  site $-1$ reaches  site $-L$ is zero, i.e., $q_{L-1}^L \to 0$ for $L\to \infty$. 
Thus, a particle never visit  site $-L$. 
This follows that the probability that a random walker visits  site $i$ $(n+1)$th times 
is given by 
\begin{widetext}
\begin{equation}
S(n) \sim  p^L_{L-i} \sum_{k=0}^n {}_nC_k (pq_1^{L-i})^k (qp_{L-1}^L)^{n-k}pp_1^{L-i}
=p^L_{L-i}  (pq_1^{L-i} + qp_{L-1}^L)^n pp_1^{L-i}
\label{Sn0}
\end{equation}
for $i\ll L$ because $q^L_{L-i}\to 0$ for $i\ll L$ , and 
\begin{equation}
S(n) \sim  q^L_{L-i} \sum_{k=0}^{n-1}  (pq_1^{L} + qp_{L-1}^L)^k pp_1^{L} (pq_1^{L-i} + qp_{L-1}^L)^{n-k-1}pp_1^{L-i}
+p^L_{L-i}  (pq_1^{L-i} + qp_{L-1}^L)^n pp_1^{L-i}
\label{Sn0}
\end{equation}
\end{widetext}
for $i\sim L$. In the large-$L$ limit, 
\begin{equation}
S(n) \to (p-q)(2q)^n\quad (L\to\infty), 
\label{Sn}
\end{equation}
for both cases $i\ll L$ and $i\sim L$. 
Therefore, the generating function of the number of visits at  site $i$ 
until the particle reaches  site $L$ is given by
\begin{equation}
Z(\lambda) = \sum_{n=0}^\infty S(n) e^{\lambda (n+1)} \to \frac{p -q}{e^{-\lambda}-2q}
\end{equation}
in the large-$L$ limit.  The generating function does not comprise $i$-dependence, that is,  the distribution of 
$k_i$ is the same for all $i$. Thus, the moments of $k_i$ do not depend on the site. In particular, 
 the mean and variance of $k_i$ are given by
\begin{equation}
\langle k_i \rangle = \left. \frac{\partial Z(\lambda)}{\partial \lambda}\right|_{\lambda =0} =\frac{1}{p-q} 
\label{k-mean}
\end{equation}
and
\begin{equation}
\langle k_i^2 \rangle - \langle k_i \rangle^2 = \left. \frac{\partial^2 \ln Z(\lambda)}{\partial^2 \lambda}\right|_{\lambda =0} 
=\frac{2q}{(p-q)^2} ,
\label{k-variance}
\end{equation}
respectively. 
Because the moments do not depend on  site $i$, we use the following notations: 
$\langle k \rangle=1/(p-q)$ and $\langle \delta k^2 \rangle =2q/(p-q)^2$. 

\section{Correlation function $\langle k_0 k_l \rangle$}

To obtain  correlation function $\langle k_0 k_l \rangle$, we consider the generating function defined by
\begin{equation}
Z(\lambda_0,\lambda_l) = \sum_{n_0,n_l} p(n_0,n_l) e^{\lambda_0 n_0}e^{\lambda_l n_l},
\end{equation}
where $p(n_0,n_l)$ is the joint probability of $k_0=n_0$ and $k_l=n_l$. Counting $k_0$ and $k_l$ for $l>1$, we have
\begin{widetext}
\begin{equation}
Z(\lambda_0,\lambda_l) = \sum_{n_0=1}^\infty \left(q + pq_1^l+p p_1^l \sum_{k=1}^\infty \{P^{(l)} \}^{k-1} e^{k \lambda_l} q q_{l-1}^l
\right)^{n_0-1} e^{\lambda_0 n_0} pp_1^l \sum_{r=0}^\infty \{P^{(l)} \}^{r} e^{\lambda_l (r+1)} p p_1^{L-l},
\end{equation}
where $P^{(l)}$ is the probability that a random walker starting from the site $l$ will return to  site $l$ without visiting 
 sites $0$ and $L$, i.e., $P^{(l)} = pq_1^{L-l} + q p_{l-1}^l$. In the large-$L$ limit, the generating function becomes
 \begin{equation}
Z(\lambda_0,\lambda_l) = \frac{e^{\lambda_0+\lambda_l} p p_1^l (p-q)}{(1-P^{(l)} e^{\lambda_l}) (1-(q+pq_1^l)e^{\lambda_0}) 
-pqq_{l-1}^l p_1^l e^{\lambda_0 + \lambda_l}},
\end{equation}
which satisfies the normalization, i.e., $Z(0,0)=1$. 

The correlation function is given by
\begin{equation}
\langle k_0 k_l \rangle = \left. \frac{\partial^2 Z}{\partial \lambda_0 \lambda_l} \right|_{\lambda_0=\lambda_l=0}
=\frac{1}{(p-q)^2} + \frac{\varepsilon^l}{(p-q)^2},
\end{equation}
where $\varepsilon \equiv q/p <1$. Thus, the correlation function decays exponentially. We note that this expression is valid 
for $l=1$ because  $q_0^1=p_1^1=1$ and $p_0^1=0$.

Next, we consider $\langle k_i k_{i+l} \rangle$. 
Because the probability that a random walker starting from the origin visits  site $-L+i+l$ for $i+l \ll L$ becomes zero in the large-$L$ limit, 
  for $i+l \ll L$ the generating function $Z(\lambda_i,\lambda_{i+l})$  is given by
\begin{equation}
Z(\lambda_i,\lambda_{i+l}) = \sum_{n_i=1}^\infty \left(q + pq_1^l+p p_1^l \sum_{k=1}^\infty \{P_i^{(l)} \}^{k-1} e^{k \lambda_{i+l}} 
q q_{l-1}^l
\right)^{n_i-1} e^{\lambda_i n_i} pp_1^l \sum_{r=0}^\infty \{P_i^{(l)} \}^{r} e^{\lambda_{i+l} (r+1)} p p_1^{L-i-l},
\end{equation}
where $P_i^{(l)} = pq_1^{L-i-l} + q p_{l-1}^l$, which is equivalent to $P^{(l)}$ in the large-$L$ limit. Moreover, $p_1^{L-i-l}=p_1^{L-l}$ in  the large-$L$ limit. Therefore, $Z(\lambda_i,\lambda_{i+l})$ is the same as $Z(\lambda_0,\lambda_l)$ in the large-$L$ limit. 
It follows that  correlation function $\langle k_i k_{i+l} \rangle$ does not depend on $i$ and is the same as $\langle k_0 k_l \rangle$.

For $i\sim L$, a random walker will visit site $-L+i+l$ with  probability $q_{L-i-l}^{L-l}$, which is nonzero even in the large-$L$ limit. 
Thus, the generating function becomes
\begin{eqnarray}
Z(\lambda_i,\lambda_{i+l}) &=& q_{L-i-l}^{L-l} \sum_{n_{i+l}=1}^\infty \left(pq_1^{L-l} + qp_{l-1}^l 
+q q_{l-1}^l \sum_{k=1}^\infty \{P_{-L+i} \}^{k-1} e^{k \lambda_i} p p_{1}^l
\right)^{n_{i+l}-1} e^{\lambda_{i+l} n_{i+l}} \nonumber\\
&\times& p p_1^{L-l} \sum_{n_i=1}^\infty \left(q + pq_1^l +p p_1^l \sum_{k=1}^\infty \{P_i^{(l)} \}^{k-1} e^{k \lambda_{i+l} }q q_{l-1}^l
\right)^{n_i-1} e^{\lambda_i n_i} pp_1^l \sum_{r=0}^\infty \{P_i^{(l)} \}^{r} e^{\lambda_{i+l} (r+1)} p p_1^{L-i-l}\nonumber\\
&+& p_{L-i-l}^{L-l} \sum_{n_i=1}^\infty \left(q + pq_1^l +p p_1^l \sum_{k=1}^\infty \{P_i^{(l)} \}^{k-1} e^{k \lambda_{i+l}} q q_{l-1}^l
\right)^{n_i-1} e^{\lambda_i n_i} pp_1^l \sum_{r=0}^\infty \{P_i^{(l)} \}^{r} e^{\lambda_{i+l} (r+1)} p p_1^{L-i-l},~~~~~
\end{eqnarray}
where $P_{-L+i} = q +pq_1^l$ and $P_i^{(l)}=pq_1^{L-i-l} + q p_{l-1}^l$. 
By a straightforward calculation, we have 
\begin{equation}
\langle k_i k_{i+l} \rangle = \left. \frac{\partial^2 Z}{\partial \lambda_i \lambda_{i+l}} \right|_{\lambda_i=\lambda_{i+l}=0}
=\frac{1}{(p-q)^2} + \frac{\varepsilon^l}{(p-q)^2}.
\end{equation}
Therefore, the correlation function does not depend on $i$.

\section{Derivation of the VFPT in the biased QTM}
The second moment of  $T_i$ can be calculated as
 $\langle T_i^2 \rangle = \langle (\tau_1^{(i)} + \cdots + \tau_{k_i}^{(i)})^2 \rangle 
 = \langle k^2 \rangle \langle (\tau_l^{(i)}) \rangle^2 + \langle k \rangle (\langle (\tau_l^{(i)})^2 \rangle -\langle (\tau_l^{(i)}) \rangle^2)$, 
 where $\langle k^2 \rangle = \langle \delta k^2 \rangle + \langle k \rangle^2$, and 
  $\langle (\tau_l^{(i)})^2 \rangle = 2\tau_i^2$ when the waiting-time distribution is the exponential distribution. 
Therefore, the second moment of the FPT in the QTM can be represented as 
\begin{eqnarray}
 \langle T^2 \rangle_L &=& \sum_{i=0}^{L-1} \langle T_i^2 \rangle +\sum_{i\ne j} \langle T_i T_j \rangle 
 = \langle k^2 \rangle  \sum_{i=0}^{L-1} \tau_i^2 
 + \langle k \rangle  \langle \delta \tau^2 \rangle_L L +\sum_{i\ne j} \langle k_i k_j \rangle \tau_i \tau_j,
 \label{T2-L-cal}
 \end{eqnarray}
 where 
 \begin{equation}
 \langle \delta \tau^2 \rangle_L = \frac{1}{L} \sum_{i=0}^{L-1} (\langle (\tau_l^{(i)})^2 \rangle -\langle (\tau_l^{(i)}) \rangle^2).
 \end{equation} 
 The third term is given by $\sum_{i\ne j} \langle k_i k_j \rangle \tau_i \tau_j = 2  \sum_{l=1}^{L-1} C_l \sum_{i=0}^{L-1-l} \tau_i \tau_{i+l}
+ \langle k \rangle^2 \sum_{i\ne j}  \tau_i \tau_j =2  \sum_{l=1}^{L-1} C_l \sum_{i=0}^{L-1-l} \tau_i \tau_{i+l}
+ \langle k \rangle^2 [ (\sum_{i=0}^{L-1}\tau_i)^2 - \sum_{i=0}^{L-1}  \tau_i^2]$, where we set $C_l \equiv \langle k_i k_{i+l}\rangle -\langle k \rangle^2$ because $\langle k_i k_{i+l}\rangle$ does not depend on $i$. Therefore, the second moment of the FPT becomes
\begin{eqnarray}
 \langle T^2 \rangle_L  &=& \langle \delta k^2 \rangle  \sum_{i=0}^{L-1} \tau_i^2 
 + \langle k \rangle  \langle \delta \tau^2 \rangle_L L + 2  \sum_{l=1}^{L-1} C_l \sum_{i=0}^{L-1-l} \tau_i \tau_{i+l} + \frac{L^2 \mu_L^2}{(p-q)^2}.
 \label{T2-L-cal2}
 \end{eqnarray}
 The combination of Eqs.~(\ref{k-mean}), (\ref{k-variance}), and (\ref{cl-asympt}) gives Eq.~(\ref{var_fpt_qtm2}).

\end{widetext}

\section{Some distribution functions}
Here, we define the probability density functions (PDFs) of $\tau_i \tau_{i+1}$ and $\tau_i^2$, where $\tau_i$ and $\tau_{i+1}$ are independent and identically distributed random variables with a power-law distribution of $\psi_\alpha(\tau) =\alpha \tau^{-1-\alpha}$ ($\tau\geq 1$). First, the probability that $\tau_i^2$ is smaller than $x$ is given by
\begin{equation}
\Pr (\tau^2 < x) = \Pr (\tau < \sqrt{x}) = 1- x^{-\alpha/2}. 
\end{equation}
Therefore, the PDF of $\tau_i^2$ is given by $\psi_{\alpha/2} (x)$. Next, the probability that $\tau_i\tau_{i+1}$ is smaller than $x$ is given by
\begin{eqnarray}
\Pr (\tau_i \tau_{i+1} \leq x) &=& \int_1^x \psi_\alpha (y) \Pr (\tau \leq x/y) dy \\
&=& 1  -  \alpha  x^{-\alpha} -  \alpha  x^{-\alpha}\ln x.
\end{eqnarray}
Thus, the PDF $\psi_{\tau_i \tau_{i+1}}(x)$ of $\tau_i \tau_{i+1}$ becomes
\begin{equation}
\psi_{\tau_i \tau_{i+1}}(x) \propto x^{-1-\alpha} \ln x \quad (x\to \infty). 
\end{equation}

\section{Asymptotes}

Here, we consider the asymptotic behavior of $\sum_{l=1}^{L-1} C_l \sum_{i=0}^{L-1-l} \tau_i \tau_{i+l}$. 
First, we show that $a_l (L)=\sum_{i=0}^{L-1-l} \tau_i \tau_{i+l}$ satisfies $a_1 \sim a_m$ for $m\ll L$ in the large-$L$ limit. 
In the large-$L$ limit, 
\begin{equation}
a_1(L) - a_m(L) \sim \sum_{i=0}^{L-m-1} \tau_i \Delta \tau_{i,m},
\end{equation}
where $\Delta \tau_{i,m} =\tau_{i+1} -\tau_{i+m}$. Because $\tau_i$ and $\Delta \tau_{i,m}$ are independent and 
$\sum _{i=0}^{L-m-1} \Delta \tau_{i,m}/L \to 0$ for $L\to \infty$, the order of $a_1(L) - a_m(L)$ is at most 
that of $\sum _{i=0}^{L-1} \tau_i $, i.e, $O(L^{1/\alpha+1})$. It follows that $a_1(L) - a_m(L) = o(a_1(L))$ because 
the PDF of $\tau_i \tau_{j}$ follows 
\begin{equation}
\psi (x) \propto x^{-1-\alpha} \ln x\quad (x\to \infty),
\end{equation}
where $\tau_i$ and $\tau_{j}$ are independent.  Therefore, we have
$a_1 \sim a_m$ for $m\ll L$ in the large-$L$ limit.

Because $C_l$  decays exponentially to zero and $a_1 \sim a_l$ for $l\ll L$,  
 $\sum_{l=1}^{L-1} C_l a_l(L)$ can be approximated 
by 
\begin{equation}
\sum_{l=1}^{L-1} C_l a_l (L) \sim  \sum_{l=1}^{m}  C_l a_l (L)\sim C  a_1 (L), 
\end{equation}
where $m$ is small and does not depend on $L$ and $C\equiv \sum_{l=1}^{L-1} C_l$. 

Next, we show $\mu_L^2 \sim a_1/L$ for $\alpha>1$ in the large-$L$ limit. 
Because the sum of $\mu_L - \tau_{i+1}$ 
is a small order of $L$, i.e.,  
\[
 \sum_{i=0}^{L-2} (\mu_L - \tau_{i+1}) \sim \tau_0,
 \]
we have 
\begin{equation}
\mu_L^2 - \frac{a_1 (L)}{L} 
\sim  \frac{1}{L} \sum_{i=0}^{L-2} \tau_i (\mu_L - \tau_{i+1}).
\end{equation}
For $\alpha>1$, $\mu_L^2 - a_1 (L)/L$ becomes small (and is a small order of $\mu_L^2$) because $\sum_{i=1}^L\tau_i \tau_{i+1}/L \cong \mu_L^2$ 
for large $L$.
 Because $\mu_L$ is a small order of $a_1/L$, we have $\mu_L^2 - a_1 (L)/L = o(a_1 (L)/L )$, i.e., 
 \begin{equation}
  a_1 \sim L \mu_L^2 \quad (L\to \infty).
  \end{equation}
For $\alpha>1$, 
\begin{equation}
\sum_{l=1}^{L-1} C_l a_l (L) \sim L \mu_L^2. 
\label{cl-asympt}
\end{equation}
 For $\alpha<1$, the generalized central limit theorem states that 
\begin{equation}
\frac{\sum_{i=1}^L \tau_i \tau_{i+1}}{L^{1/\alpha}} \Rightarrow Y_\alpha,
\end{equation}
where $\Rightarrow$ implies the convergence in distribution and $Y_\alpha$ is a random variable with a stable distribution with 
 index $\alpha$ and
\begin{equation}
\frac{\sum_{i=1}^L \tau_i^2}{L^{2/\alpha}} \Rightarrow Y_{\alpha/2}.
\end{equation}
Thus, $a_1(L)/\sigma_L^2 \to 0$ for $L\to \infty$. Therefore, 
both $\mu_L^2$ and $a_l(L)$ are  small orders of $\sigma_L^2$. Thus, these terms  in 
Eq.~(\ref{T2-L-cal}) can be ignored.

\section{Disorder average and correlation plot for the MFPT and  VFPT in the infinite 1D systems with nonperiodic landscapes}

\begin{figure}
\includegraphics[width=.9\linewidth, angle=0]{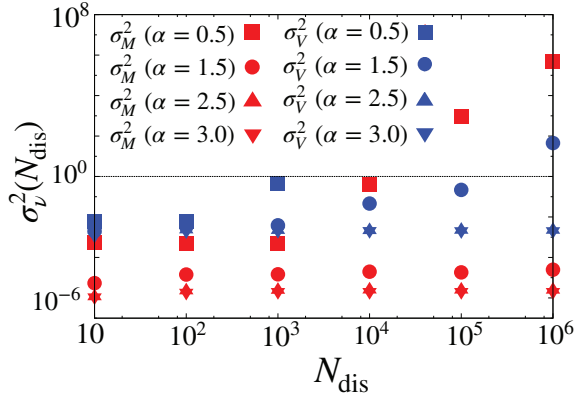}
\caption{Variances of $T_{\rm M}$ and $T_{\rm V}$ as a function of $N_{\rm dis}$ ($L = 10^3$ and $p = 0.8$). Symbols represent results of numerical simulations. We used $10^3$ thermal paths to calculate the MFPT or VFPT for a fixed disorder realization.}
\label{variance-infinite}
\end{figure}

 To quantify the degree of the disorder average, we introduce a sample-number-dependent variance:
\begin{equation}
\sigma_\nu^2( N_{\rm dis}) \equiv  \sum_{j=1}^{N_{\rm dis}} \frac{ (T_{\nu}^{(j)} )^2}{N_{\rm dis}} - \left(\sum_{j=1}^{N_{\rm dis}} \frac{T_{\nu}^{(j)}}{N_{\rm dis}}\right)^2 \, ,   ~~~~{\nu={\rm M}, {\rm V}} 
 \end{equation}
where $T_{\nu}^{(j)}$ is a numerical value of $T_{\nu}$ for the $j$th realization of disorder. This quantifies sample-to-sample fluctuations of $T_{\nu}^{(j)}$   
as a function of $N_{\rm dis}$. With increasing $N_{\rm dis}$,  the average approaches the exact disorder average. From the indication of the CTRW results (\ref{mfptvar_fpt_ctrw}) and (\ref{vfptvar_fpt_ctrw}), the exact disorder averages of FPT statistics will diverge  for small $\alpha$.
In Fig.~\ref{variance-infinite}, we show the $N_{\rm dis}$-dependence of $\sigma_\nu^2( N_{\rm dis})$.
The figure clearly shows that  $\sigma_{\rm M}^2( N_{\rm dis}) $ and $\sigma_{\rm V}^2( N_{\rm dis}) $ become divergent behaviors for $\alpha < 1$ and $\alpha <2$, respectively, as increasing $N_{\rm dis}$.

\begin{figure}
\includegraphics[width=.95\linewidth, angle=0]{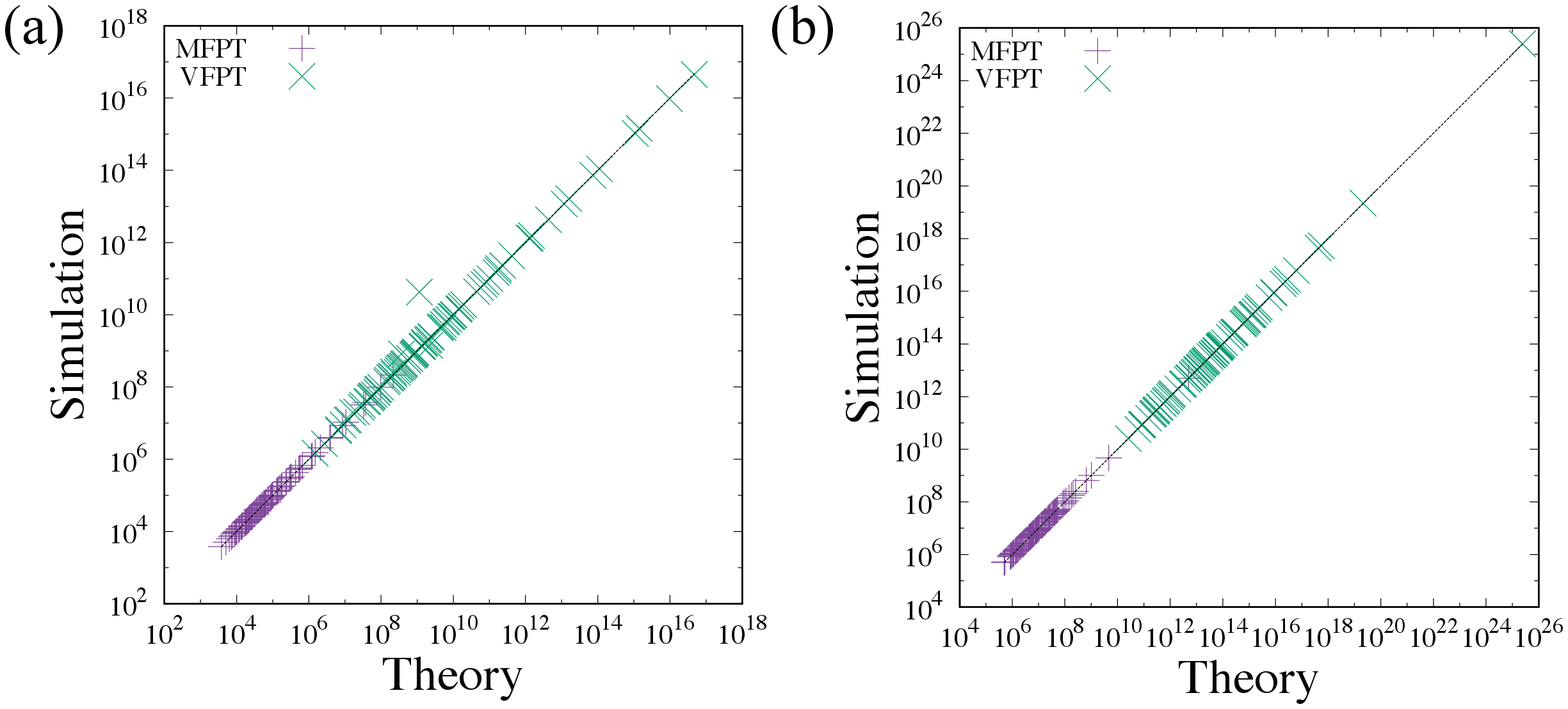}
\caption{Correlation plot for the MFPT and  VFPT in the infinite 1D systems with nonperiodic landscapes ($\alpha=0.5$) 
for (a) $L=10^2$ and (b) $L=10^3$. 
Numerical simulations of the MFPT and  VFPT for 100 disorder realizations  are presented by symbols. Some points deviate 
from the line, indicating the theories do not work well in these disorder realizations.
}
\label{cp-infinite}
\end{figure}

Figure~\ref{cp-infinite} presents numerical data of $\langle T \rangle_L$ and $\langle \delta T^2 \rangle_L$. Numerical results on the $y=x$ line imply that 
$T_{\rm MFPT}'$ and $T_{\rm VFPT}'$ are very good approximations of the MFPT and VFPT, respectively. However, there  are a few realizations that deviate from the 
$y=x$ line.  This figure explicitly explains a mechanism of the divergence according to the disorder average; i.e., the divergence is caused by a small proportion of samples with extremely large deviations. In such rare samples, significantly large waiting times are assigned for $i<0$ and small $|i|$. 
 Except for such rare samples, the biased-QTM results with a periodic landscape are surprisingly useful 
in nonperiodic landscapes.

\section{Derivation of the CTRW results}
We derive the MFPT and  VFPT in the CTRW following Ref.~\cite{Condfamin2007}, and use the same notations as in \cite{Condfamin2007} to 
 present the MFPT as follows:
\begin{equation}
\langle T_{\rm \tiny ctrw} \rangle_{L} = \int_0^\infty t \pi(t) dt = \sum_{n=1}^\infty Q(n) \int_0^\infty t \psi_n(t)dt .
\end{equation}
The Laplace transform of $\psi_n(t)$ can be given by $\hat{\psi}_n(s) = \hat{\psi}(s)^n$. It follows that 
\begin{equation}
\int_0^\infty t \psi_n(t)dt = -n \hat{\psi}(0)^{n-1} \hat{\psi}'(0) =n\mu.
\end{equation}
Based on the classical RW result, we obtain
\begin{equation}
\langle T_{\rm \tiny ctrw} \rangle_L  = \mu \sum_{n=1}^\infty nQ(n) =\frac{\mu L}{p-q}.
\end{equation}
Similarly, we have 
\begin{eqnarray}
\langle T_{\rm \tiny ctrw}^2 \rangle_{L} &=& \sum_{n=1}^\infty Q(n) \int_0^\infty t^2 \psi_n(t)dt \\
&=& \sum_{n=1}^\infty Q(n) (n^2 \sigma^2 + n \mu^2)\\
&=& \mu^2\left( \frac{4pqL}{(p-q)^3} + \frac{L^2}{(p-q)^2} \right) + \frac{\sigma^2 L}{p-q}. ~~~
\end{eqnarray}
Thus, the VFPT of the biased CTRW becomes Eq.~(\ref{mfptvar_fpt_ctrw}).

\end{document}